\documentstyle[aps,prb,eqsecnum,epsf,floats]{revtex}
\begin{document}
\def\SNG{{\em Physical Review Style and Notation Guide}}
\def\LUG {{\em \LaTeX{} User's Guide \& Reference Manual}}
\def\btt#1{{\tt$\backslash$\string#1}}%
\def\REVTeX{REV\TeX}
\def\AmS{{\protect\the\textfont2
        A\kern-.1667em\lower.5ex\hbox{M}\kern-.125emS}}
\def\AmSLaTeX{\AmS-\LaTeX}
\def\BibTeX{\rm B{\sc ib}\TeX}
\twocolumn[\hsize\textwidth\columnwidth\hsize\csname@twocolumnfalse%
\endcsname
 
\title{Nonanalytic behavior of the spin susceptibility in clean Fermi systems}
\author{D.Belitz}
\address{Department of Physics and Materials Science Institute,
University of Oregon,
Eugene, OR 97403}
\author{T.R.Kirkpatrick}
\address{Institute for Physical Science and Technology, 
         and Department of Physics\\
         University of Maryland, College Park, MD 20742}
\author{Thomas Vojta}
\address{Department of Physics and Materials Science Institute,
                          University of Oregon, Eugene, OR 97403\\
        and Institut f{\"u}r Physik, Technische Universit{\"a}t
            Chemnitz-Zwickau, D-09107 Chemnitz, FRG}
\date{\today}
\maketitle
 
\begin{abstract}
The wavevector and temperature dependent static spin susceptibility, 
$\chi_s({\bf Q},T)$, of clean interacting Fermi systems is considered in 
dimensions $1\leq d \leq 3$. We show that at zero temperature $\chi_s$ 
is a nonanalytic 
function of $\vert {\bf Q}\vert$, with the leading nonanalyticity being 
$\vert{\bf Q}\vert^{d-1}$ for $1<d<3$, and ${\bf Q}^2\ln\vert{\bf Q}\vert$ 
for $d=3$. For the homogeneous spin susceptibility we find a nonanalytic
temperature dependence $T^{d-1}$ for $1<d<3$. We give qualitative mode-mode 
coupling arguments
to that effect, and corroborate these arguments by a perturbative
calculation to second order in the electron-electron interaction amplitude.
The implications of this, in particular for itinerant ferromagnetism, are
discussed. We also point out the relation between our findings and established
perturbative results for $1$-$d$ systems, as well as for the temperature 
dependence of $\chi_s({\bf Q}=0)$ in $d=3$.
\end{abstract}

]
\section{Introduction}
\label{sec:I}

It is well known that in fluids, that is in interacting many-body systems,
there are long-range correlations between the particles. 
For example, in classical fluids in
thermal equilibrium there are dynamical long-range correlations that
manifest themselves as long-time tails, or power-law decay of equilibrium
time correlation functions at large times.\cite{DKS,ernst} In frequency space,
the analogous effects are nonanalyticities at zero frequency. In an intuitive
physical picture, these correlations can be understood as memory effects:
The particles ``remember'' previous collisions, and therefore so-called
ring collision events, where after a collision the two involved particles
move away and later recollide, play a special role for the dynamics of the
fluid. Technically, the long-time tails can be described in terms of mode-mode
coupling theories. The salient point is that with any quantities whose
correlations constitute soft, or gapless, modes (due to conservation laws,
or for other reasons), products of these quantities have the same 
property.\cite{Forster}
In the equations of motion that govern the behavior of time correlation
functions this leads to convolutions of soft propagators, which in turn
results in nonanalytic frequency dependences. For phase space reasons, the
strength of the effect increases with decreasing dimensionality: While in
three-dimensional ($3$-$d$) classical fluids the long-time tails provide
just a correction to the asymptotic hydrodynamic description of the system,
in $2$-$d$ fluids they are strong enough to destroy 
hydrodynamics.\cite{DKS,ForsterNelsonStephen}
 
A natural question to ask is whether such long-range correlations also occur
in position space. Indeed, in classical fluids in nonequilibrium steady states
effects occur that may be considered as the spatial analogs of long-time
tails, but in thermal equilibrium this is not the case.\cite{DKS,ernst} 
This changes, however, if we consider quantum fluids.
The quantum nature of a system has two major implications as far as 
statistical mechanics is concerned. First, temperature enters, apart from
occupation numbers, through Matsubara frequencies, which means that the
system's behavior as a function of temperature will in general be the same 
as its behavior as a function of frequency, at least at asymptotically low
temperatures. Second, and more importantly, in quantum statistical mechanics
statics and dynamics are coupled and need to be considered together. This
raises the question of whether in a quantum fluid there might be
long-range spatial correlations even in equilibrium.

From studies of systems with quenched disorder, there is evidence for the
answer to this question being affirmative. Let us consider interacting
fermions in an environment of static scatterers. In dimensions
$d>2$, and for a sufficiently small scatterer density, the relevant soft
modes in such a system are diffusive, so frequency $\Omega$, or temperature
$T$, scales like the square of the wavevector ${\bf Q}$, 
$\Omega \sim T \sim {\bf Q}^2$. 
Via mode-mode coupling effects that are analogous
to those present in classical fluids, dynamical long-range correlations
lead to long-time tails in equilibrium time correlation functions. For
instance, the electrical conductivity
as a function of frequency behaves like $\Omega^{(d-2)/2}$ at
small $\Omega$ in $d>2$ dimensions.\cite{R} The dynamical spin susceptibility,
$\chi_s({\bf Q},\Omega)$, shows no analogous long-time tail at ${\bf Q}=0$
for reasons related to spin conservation. However, from the above 
arguments about the coupling of statics and dynamics in quantum statistical
mechanics and the scaling of frequency with wavenumber, one would
expect the {\em static} spin susceptibility, $\chi_s({\bf Q},\Omega=0)$
at $T=0$, to show a related nonanalyticity at ${\bf Q}=0$, namely
$\chi_s \sim \vert {\bf Q}\vert^{d-2}$. This is indeed the case, 
as can be seen most easily
from perturbative calculations.\cite{fm_dirty} Schematically, the coupling
of two diffusive modes leads to contributions to $\chi_s$ of the type,
\begin{equation}
\int d{\bf q} \int d\omega\ \frac{1}{\omega + {\bf q}^2}\ 
   \frac{1}{\omega + \Omega + ({\bf q} + {\bf Q})^2}\quad,
\label{eq:1.1}
\end{equation}
which leads to the above behavior. One can then invoke renormalization
group arguments to show that this is indeed the leading small-${\bf Q}$ 
behavior of $\chi_s$. Similarly, at finite temperature the homogeneous
susceptibility behaves like $\chi_s({\bf Q}=0,\Omega=0) \sim T^{(d-2)/2}$.
This has interesting consequences for itinerant 
magnetism in such systems, as has been recently 
discussed.\cite{fm_dirty,fm_dirty_epl,fm_sb}

Somewhat surprisingly, the situation is much less clear in clean Fermi
systems. Here the soft modes are density and spin density fluctuations,
as well as more general particle-hole excitations. All of these have
a linear dispersion relation, i.e. $\Omega \sim \vert {\bf Q}\vert$. 
The form of the dispersion relation does not affect the basic physical 
arguments for
nonanalytic frequency and wavenumber dependences given above. One might
thus expect the spin susceptibility to have mode-mode coupling contributions
of a type analogous to those shown in Eq.\ (\ref{eq:1.1}), but with
ballistic instead of diffusive modes:
\begin{equation}
\int d{\bf q} \int d\omega\ \frac{1}{\omega + \vert{\bf q}\vert}\ 
                  \frac{1}{\omega + \Omega + \vert{\bf q}+{\bf Q}\vert}\quad,
\label{eq:1.2}
\end{equation}
which leads to $\chi_s({\bf Q},\Omega=0)\sim {\rm const} + 
\vert {\bf Q}\vert^{d-1}$ in generic
dimensions at $T=0$. In $d=3$, one would expect a 
${\bf Q}^2\ln\vert {\bf Q}\vert$ behavior, as
convolution integrals tend to yield logarithms in special dimensions.
Such a behavior of $\chi_s$ would have profound consequences for the
critical behavior of itinerant ferromagnets, as has been pointed out
recently.\cite{fm_clean} It is therefore of importance to unambigiously
determine whether or not the above mode-mode coupling arguments do indeed
carry over from disordered to clean systems.

Before we start this task, let us discuss the available information concerning
long-range correlations in clean Fermi systems. The specific heat is known to
be a nonanalytic function of temperature, viz. $C_V/T \sim T^2\ln T$ in
$d=3$. This is a consequence of a nonanalytic correction to the linear
dispersion relation of the quasiparticles in Fermi liquid theory, namely
$\Delta\epsilon (p) \sim (p-p_F)^3\ln\vert p-p_F\vert$.\cite{BaymPethick} 
Such a nonanalyticity signalizes the presence of a long-range effective
interaction between the quasiparticles, and in general it will lead to
nonanalytic behavior of both thermodynamic quantities and time correlation
functions. The $T^2\ln T$ term in the specific heat coefficient is an
example of such an effect. In $d=2$ the behavior is 
$C_V/T \sim T$,\cite{CoffeyBedell} which is
consistent with the behavior $C_V/T \sim T^{d-1}$ in generic dimensions
that one would expect from the above arguments. 
It was natural to look for similar effects
in other quantities, in particular in the spin susceptibility. These
investigations concentrated on the temperature dependence of $\chi_s$, and
several authors indeed reported to have found a $T^2\ln T$ term in the
homogeneous static spin susceptibility. However, other investigations did
not find such a contribution.\cite{controversy} The resulting confusion
has been discussed by Carneiro and Pethick.\cite{CarneiroPethick} These
authors used Fermi liquid theory to show that, while $T^2\ln T$ terms
do indeed appear in intermediate stages of the calculation of $\chi_s$ as
well as of $C_V$, they cancel in the former.

This somewhat surprising result casts some doubt on the general physical
picture painted above, which suggests the qualitative equivalence of disordered
and clean systems with respect to the presence of long-range
correlations, and resulting nonanalyticities in both the statics and
the dynamics of quantum systems. On the other hand, a failure of this
general picture would be hard to understand from several points of view.
For instance, in $d=1$ the instability of the Fermi liquid with respect
to the Luttinger liquid is well known to manifest itself in perturbation
theory for $\chi_s$ by means of logarithmic 
singularities.\cite{DzyaloLarkin,Solyom} This is precisely what one
obtains from the mode-mode coupling integral, Eq.\ (\ref{eq:1.2}). By
continuity one therefore expects $\chi_s({\bf Q}=0,T) \sim T^{d-1}$, 
and $\chi_s({\bf Q},T=0) \sim \vert {\bf Q}\vert^{d-1}$,
at least in $d=1+\epsilon$. Unless the physics changes qualitatively between 
$d=1+\epsilon$ and $d=3$, this should still be true in higher dimensions. 
Also, the corrections to Landau theory we are discussing here can be cast in
the language of the renormalization group. In this
framework, the Fermi liquid ground state is described as a stable
fixed point,\cite{Shankar} and
the effects we are interested in manifest themselves as
an irrelevant operator that leads to corrections to scaling near this
fixed point.\cite{ustbp}
In a system where ${\bf Q}$, $\Omega$, and $T$ all have a
scale dimension of unity, this operator should appear as 
$\vert {\bf Q}\vert^{d-1}$, $\Omega^{d-1}$, etc., dependences 
in various correlation functions. From a general scaling point of view
it would be hard to understand if this were not the case, except for
the possibility that the prefactors of some nonanalyticities might
accidentally vanish in certain dimensions.

It is the purpose of the present paper to clarify this confusing
point. We will show that the above general physical picture does indeed
hold true, and that it is not violated by the previously found absence
of a $T^2\ln T$ term in $\chi_s$ in $d=3$, which is accidental. The 
remainder of this paper is organized as follows.
In Sec.\ \ref{sec:II} we define our model.
In Sec.\ \ref{sec:III} we perform an explicit perturbative
calculation to second order in the electron-electron interaction. This
confirms both our qualitative arguments, and the results of 
Ref.\ \onlinecite{CarneiroPethick}. We explain why there is no
contradiction between these results, and we also make contact with
established perturbative results in $d=1$. In Sec.\ \ref{subsec:IV.A}
we discuss our result in the light of mode-mode coupling arguments that
are an elaboration of those given above. In
Sec.\ \ref{subsec:IV.B} we make contact with renormalization group ideas,
and argue that the functional forms of the nonanalyticities derived in
Sec.\ \ref{sec:III} by means of perturbation theory are asymptotically
exact. In Sec.\ \ref{subsec:IV.C} we discuss the physical consequences of
our results.

\section{Model, and Theoretical Framework}
\label{sec:II}

\subsection{The Model}
\label{subsec:II.A}

Let us consider a system of clean fermions governed by an
action\cite{fieldtheoryfootnote}
\begin{mathletters}
\label{eqs:2.1}
\begin{equation}
S = -\int dx \sum_{\sigma} \bar\psi_{\sigma}(x)\, \frac{\partial}{\partial\tau}
     \,\psi_{\sigma}(x) + S_0 + S_{\rm int}\quad.
\label{eq:2.1a}
\end{equation}
Here we use a four-vector notation, $x\equiv ({\bf x},\tau)$, and
$\int dx \equiv \int d{\bf x} \int_{0}^{\beta} d\tau$. 
${\bf x}$ denotes position,
$\tau$ imaginary time, $\beta = 1/T$, and we choose units such that
$\hbar = k_B = 1$. $\sigma$ is the spin label. 
$S_0$ describes free fermions with chemical potential $\mu$,
\begin{equation}
S_0 = \int dx \sum_{\sigma} \bar\psi_{\sigma}(x)\left[\Delta/2m + \mu\right]
      \psi_{\sigma}(x)\quad,
\label{eq:2.1b}
\end{equation}
with $\Delta$ the Laplace operator, and $m$ the fermion mass. $S_{\rm int}$
describes a two-particle, spin independent interaction,
\begin{eqnarray}
S_{\rm int} &=& -\frac{1}{2}\int dx_1\,dx_2 \sum_{\sigma_1,\sigma_2} v(x_1-x_2)
\nonumber\\
   &&\times\bar\psi_{\sigma_1}(x_1)\bar\psi_{\sigma_2}(x_2) 
     \psi_{\sigma_2}(x_2)\psi_{\sigma_1}(x_1)\quad.
\label{eq:2.1c}
\end{eqnarray}
\end{mathletters}%
The interaction potential $v(x)$ will be specified in Sec.\ \ref{subsec:II.B}
below.

\begin{figure}
\epsfxsize=8.0cm
\epsfysize=8.0cm
\epsffile{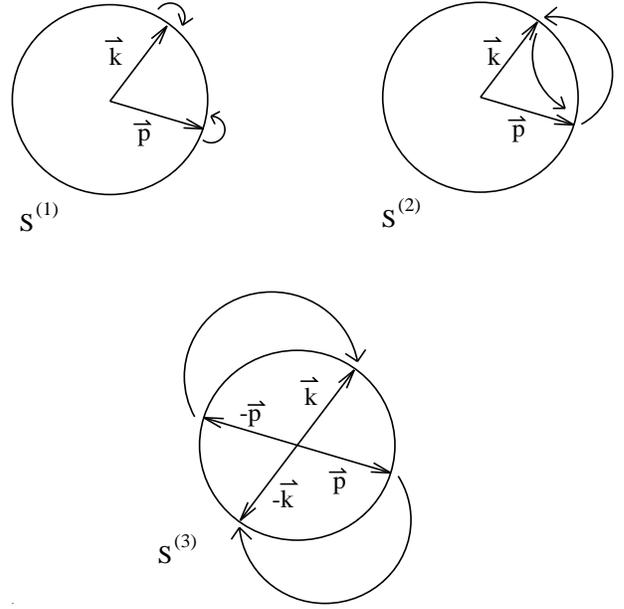}
\vskip 0.5cm
\caption{Typical small-angle (1), large-angle (2), and $2k_F$-scattering
 processes (3) near the Fermi surface in $d=2$.}
\label{fig:1}
\end{figure}

We now Fourier transform to wave vectors ${\bf k}$ and 
fermionic Matsubara frequencies
$\omega_n = 2\pi T(n+1/2)$. Later we will also encounter bosonic Matsubara
frequencies, which we denote by $\Omega_n = 2\pi Tn$. 
Using again a four-vector notation, $k\equiv ({\bf k},\omega_n)$, 
$\sum_k \equiv T\sum_{i\omega_n}\ \int d{\bf k}/(2\pi)^d$,
we can write,
\begin{mathletters}
\label{eqs:2.2}
\begin{equation}
S_0 = \sum_{\sigma} \sum_{k} \bar\psi_{\sigma}(k) \left[i\omega_n - {\bf k}^2/2m
      + \mu\right]\psi_{\sigma}(k)\quad,
\label{eq:2.2a}
\end{equation}
\begin{eqnarray}
S_{\rm int} &=& \frac{-T}{2} \sum_{\sigma_1,\sigma_2} \sum_{\{k_i\}}
              \delta_{k_1+k_2,k_3+k_4}\ v(k_2-k_3)
\nonumber\\
      &&\times\bar\psi_{\sigma_1}(k_1)\bar\psi_{\sigma_2}(k_2)
                \psi_{\sigma_2}(k_3)\psi_{\sigma_1}(k_4)\quad.
\label{eq:2.2b}
\end{eqnarray}
\end{mathletters}%
For the long-wavelength, low-frequency processes we will be interested
in, only the scattering of particles and holes close to the Fermi surface
is important. It is customary and convenient to divide these processes
into three classes:\cite{AGD} (1) Small-angle scattering, 
(2) large-angle scattering, and (3) $2k_F$-scattering. These classes are
also referred to as the particle-hole channel for classes (1) and (2), and
the particle-particle or Cooper channel for class (3), respectively. The
corresponding scattering processes are
schematically depicted in Fig.\ \ref{fig:1}.
For our purposes it is convenient to make the phase space decomposition
that is inherent to this classification explicit by writing the interaction
part of the action,
\begin{mathletters}
\label{eqs:2.3}
\begin{equation}
S_{\rm int} = S_{\rm int}^{\,(1)} + S_{\rm int}^{\,(2)} + S_{\rm int}^{\,(3)}
                 \quad,
\label{eq:2.3a}
\end{equation}
where
\begin{eqnarray}
S_{\rm int}^{\,(1)} &=& \frac{-T}{2} \sum_{\sigma_1,\sigma_2} \sum_{k,p} 
                    {\sum_{q}}^{\prime} v({\bf q})\ \bar\psi_{\sigma_1}(k)
                    \bar\psi_{\sigma_2}(p+q)
\nonumber\\
                  &&\times\psi_{\sigma_2}(p)\psi_{\sigma_1}(k+q)\quad,
\label{eq:2.3b}\\
S_{\rm int}^{\,(2)} &=& \frac{-T}{2} \sum_{\sigma_1,\sigma_2} \sum_{k,p}
                  {\sum_{q}}^{\prime}v({\bf p}-{\bf k})\ \bar\psi_{\sigma_1}(k)
                  \bar\psi_{\sigma_2}(p+q)
\nonumber\\
                  &&\times\psi_{\sigma_2}(k+q)\psi_{\sigma_1}(p)\quad,
\label{eq:2.3c}\\
S_{\rm int}^{\,(3)} &=& \frac{-T}{2} \sum_{\sigma_1\neq\sigma_2} \sum_{k,p}
                  {\sum_{q}}^{\prime}v({\bf k}+{\bf p})\ \bar\psi_{\sigma_1}(k)
                  \bar\psi_{\sigma_2}(-k+q)
\nonumber\\
                  &&\times\psi_{\sigma_2}(p+q)\psi_{\sigma_1}(-p)\quad.
\label{eq:2.3d}
\end{eqnarray}
\end{mathletters}%
Here the prime on the $q$-summation indicates that only momenta up to some
cutoff momentum $\Lambda$ are integrated over. This restriction is necessary to
avoid double counting, since each of the three expressions, 
Eqs.\ (\ref{eq:2.3b}) - (\ref{eq:2.3d}), represents all of $S_{\rm int}$ if
all wavevectors are summed over. The long-wavelength physics we
are interested in will not depend on $\Lambda$.

\begin{figure}
\epsfxsize=8.0cm
\epsfysize=4.0cm
\epsffile{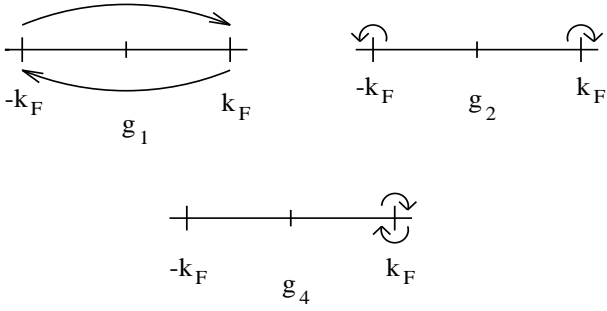}
\vskip 0.5cm
\caption{The three independent scattering processes near the Fermi surface
  with interaction amplitudes $g_1$, $g_2$, and $g_4$ in $d=1$.}
\label{fig:2}
\end{figure}

The above phase space decomposition is correct
in dimensions $d\geq 2$. In $d=1$, the Fermi surfaces collapses onto two
Fermi points, and the processes we called above large-angle scattering and
$2k_F$-scattering become indistinguishable. The three independent
scattering processes are usually chosen as the ones shown in
Fig.\ \ref{fig:2}, and the corresponding coupling interaction potentials
are denoted by $g_1$, $g_2$, and $g_4$.\cite{Solyom}
Inspection shows that the action written in Eqs.\ (\ref{eqs:2.3}) counts
each of these processes twice. If $S_{\rm int}^{\,(3)}$ is dropped,
then the $g_4$-process is still counted twice. However, it is known
that $g_4$ does not contribute to the logarithmic terms we are interested 
in.\cite{Solyom} For our purposes it therefore
is sufficient to just drop the particle-particle channel when we are
dealing with $d=1$.

\subsection{Simplifications of the Model}
\label{subsec:II.B}

The effective interaction potentials that appear in
Eqs.\ (\ref{eq:2.3b}) - (\ref{eq:2.3d}) are all given by the basic
potential $v$, taken at different momenta. $S_{\rm int}^{\,(1)}$ contains
the direct scattering contribution, or $v({\bf q})$, with ${\bf q}$ the
restricted momentum. If $v$ is chosen to be a bare Coulomb interaction,
then this leads to singularities in perturbation theory in $v$ that
indicate the need for infinite resummations to incorporate screening.
For simplicity, we assume that this procedure has already been carried
out, and take $v$ to be a statically screened Coulomb interaction. For
effects that arise from small values of $\vert{\bf q}\vert$ it is then
sufficient to replace $v({\bf q})$ by the number
$\Gamma_1 \equiv v({\bf q}\rightarrow 0)$.\cite{jellium} 
In Eqs.\ (\ref{eq:2.3c}), (\ref{eq:2.3d})
the moduli of ${\bf k}$ and ${\bf p}$ are equal to $k_F$ for the dominant
scattering processes, and one usually expands these coupling constants
in Legendre polynomials on the Fermi surface. While all of the terms in
this expansion contribute to the processes we want to study, we note
that the coefficients in the angular momentum expansion are independent
coupling constants. In order to establish the existence of a nonanalytic
term in $\chi_s({\bf Q})$, 
it therefore is sufficient to establish its existence
in a particular angular momentum channel. For simplicity we choose the
zero angular momentum channel, $l=0$. We then have three coupling constants
in our theory, namely $\Gamma_1$, and $\Gamma_2$ and $\Gamma_3$, which
are $v({\bf k}-{\bf p})$ and $v({\bf k}+{\bf p})$, respectively, averaged
over the Fermi surface. Instead of $\Gamma_1$ and $\Gamma_2$ one often uses
the particle-hole spin singlet and spin triplet interaction amplitudes
$\Gamma_s$ and $\Gamma_t$ that are linear combinations of
$\Gamma_1$ and $\Gamma_2$. They are related to the Fermi liquid parameters
$F_0^s$ and $F_0^a$ by,
\begin{mathletters}
\label{eqs:2.4}
\begin{equation}
\Gamma_s = \Gamma_1 - \Gamma_2/2 = \frac{1}{2N_F}\frac{F_0^s}{1+F_0^s}\quad,
\label{eq:2.4a}
\end{equation}
\begin{equation}
\Gamma_t = \Gamma_2/2 = \frac{-1}{2N_F}\frac{F_0^a}{1+F_0^a}\quad,
\label{eq:2.4b}
\end{equation}
where $N_F$ is the density of states at the Fermi level. Our simplified
model is tantamount to taking only $F_0^s$ and $F_0^a$ into account
instead of the complete sets of Landau parameters. As explained above,
this is sufficient for our purposes. We also define the Cooper channel
amplitude,
\begin{equation}
\Gamma_c = \Gamma_3/2\quad,
\label{eq:2.4c}
\end{equation}
\end{mathletters}%
and again we keep only the $l=0$ channel.
The particle-particle channel is neglected in Landau theory.

\begin{figure}
\epsfxsize=8.0cm
\epsfysize=9.0cm
\epsffile{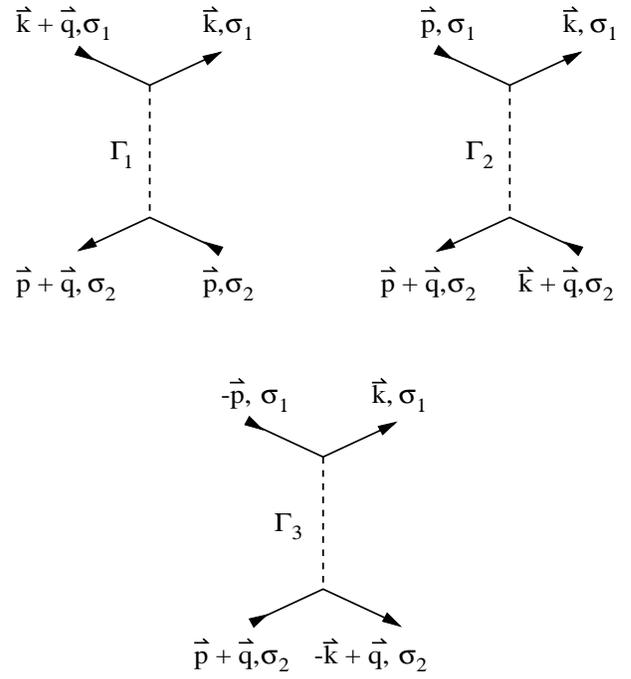}
\vskip 0.5cm
\caption{The three interaction vertices with coupling constants
 $\Gamma_1$, $\Gamma_2$, and $\Gamma_3$.}
\label{fig:3}
\end{figure}

Our model is now defined as Eqs.\ (\ref{eqs:2.2}) and (\ref{eqs:2.3}),
with $v({\bf q})$, $v({\bf p}-{\bf k})$, and $v({\bf k}+{\bf p})$
replaced by $\Gamma_1$, $\Gamma_2$, and $\Gamma_3$, respectively.
We thus have three different interaction vertices that are shown in
Fig.\ \ref{fig:3}. In the following section we will calculate $\chi_s$
in perturbation theory with respect to the interaction amplitudes
$\Gamma_1$, $\Gamma_2$, and $\Gamma_3$.

\section{Perturbation Theory}
\label{sec:III}

\subsection{Contributions to second order in the interaction}
\label{subsec:III.A}

\begin{figure}
\epsfxsize=8.0cm
\epsfysize=8.0cm
\epsffile{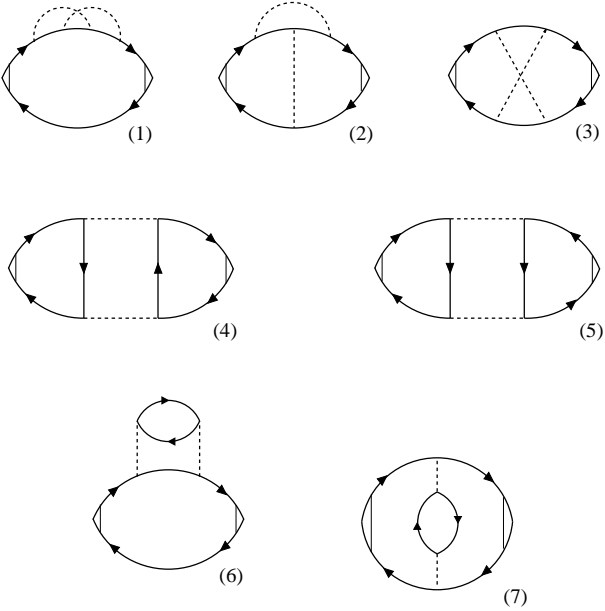}
\vskip 0.5cm
\caption{Second order diagrams that contribute to the nonanalytic behavior
 of $\chi_s$. The solid vertical line denotes the external spin vertex
 $\sigma$.}
\label{fig:4}
\end{figure}

We now proceed to calculate the spin susceptibility, $\chi_s$, in perturbation
theory with respect to the electron-electron interaction. This can be done
by means of standard methods.\cite{NegeleOrland,FetterWalecka,AGD} We will be
interested only in contributions that lead to a nonanalytic wavenumber
dependence. It is easy to see that no nonanalytic
behavior can occur at first order in the interaction. At second order,
there is also a large number of diagrams for which this is true, and others
vanish due to charge neutrality.\cite{jellium}
There remain seven topologically different second order diagrams, 
all shown in Fig.\ \ref{fig:4}, that need to be considered.
We thus write
\begin{equation}
\chi_s(Q) = 2\chi_0(Q) + \sum_{i=1}^{7}\chi^{(i)}
          +\ ({\rm analytic\ contributions})\quad,
\label{eq:3.1}
\end{equation}
where $\chi_0$ denotes the Lindhard function, and the correction terms are
labeled according to the diagrams in Fig.\ \ref{fig:4}. Here and in the
remainder of this section we use again the four-vector notation of
Sec.\ \ref{sec:II}, so $Q\equiv ({\bf Q},\Omega_n)$, etc.

These diagrams can be expressed in terms of integrals over electronic
Green's functions, or bare electron propagators, that can be read off
Eq.\ (\ref{eq:2.2a}),
\begin{equation}
G_k \equiv G_{\bf k}(i\omega_n) = \frac{1}{i\omega_n - {\bf k}^2/2m + \mu}\quad.
\label{eq:3.2}
\end{equation}
In terms of the $G_k$, we find
\begin{mathletters}
\label{eqs:3.3}
\begin{equation}
\chi^{(1)} = -4\Gamma_1\,\Gamma_2\sum_{\sigma} \sigma^2 {\sum_q}^{\prime}
             J^{(4)}_1(q,Q)\,J^{(2)}(q)\quad,
\label{eq:3.3a}
\end{equation}
\begin{eqnarray}
\chi^{(2)} &=& -2\Gamma_1\,\Gamma_2\sum_{\sigma} \sigma^2 {\sum_q}^{\prime}
             \biggl[\left( J^{(3)}(q,Q)\right)^2
\nonumber\\
           && + J^{(4)}_2(q,Q)\,J^{(2)}(q)
             \biggr]\quad,
\label{eq:3.3b}
\end{eqnarray}
\begin{equation}
\chi^{(3)} = -2\Gamma_1\,\Gamma_2\sum_{\sigma} \sigma^2 {\sum_q}^{\prime}
             J^{(3)}(q,Q)\,J^{(3)}(-q,-Q)\quad,
\label{eq:3.3c}
\end{equation}
\begin{equation}
\chi^{(4)} = (\Gamma_3)^2\sum_{\sigma_1,\sigma_2} \sigma_1 \sigma_2 
    \left(1-\delta_{\sigma_1 \sigma_2}\right) {\sum_q}^{\prime}
    I^{(3)}_1(q,Q)\,I^{(3)}_2(q,Q),
\label{eq:3.3d}
\end{equation}
\begin{equation}
\chi^{(5)} = (\Gamma_3)^2\sum_{\sigma_1,\sigma_2} \sigma_1 \sigma_2
    \left(1-\delta_{\sigma_1 \sigma_2}\right) {\sum_q}^{\prime}
    I^{(4)}_2(q,Q)\,I^{(2)}(q)\ ,
\label{3.3e}
\end{equation}
\begin{eqnarray}
\chi^{(6)} &=& 2\sum_{\sigma_1,\sigma_2} \sigma_1^2 {\sum_q}^{\prime}
             \left[(\Gamma_1)^2\,J^{(4)}_1(q,Q)
                  +(\Gamma_2)^2\,J^{(4)}_1(-q,Q)\right]
\nonumber\\
          &&\times J^{(2)}(-q) +2(\Gamma_3)^2\sum_{\sigma_1,\sigma_2} 
                                                       \sigma_1 \sigma_2
       \left(1-\delta_{\sigma_1 \sigma_2}\right) 
\nonumber\\
          &&\times {\sum_q}^{\prime} I^{(4)}_1(q,Q)\,I^{(2)}(q)\quad,
\label{eq:3.3f}
\end{eqnarray}
\begin{eqnarray}
\chi^{(7)} &=& \sum_{\sigma_1,\sigma_2} \sigma_1^2 {\sum_q}^{\prime}
             \biggl[(\Gamma_1)^2\,J^{(4)}_2(q,Q)\,J^{(2)}(q)
\nonumber\\
          &&+(\Gamma_2)^2\,\left(J^{(3)}(-q,Q)\right)^2\biggr]
            -\ \chi^{(4)}\quad.
\label{eq:3.3g}
\end{eqnarray}
\end{mathletters}%
Here $q$ is a bosonic frequency-momentum integration 
variable. In Eqs.\ (\ref{eqs:3.3}), the
following multiplication factors have been taken into account.
In diagram (1), either one of the interaction lines can be a $\Gamma_1$; the
other one is then necessarily a $\Gamma_2$. This leads to a multiplication
factor of $2$, and another factor of $2$ comes from the existence of an
equivalent symmetric diagram. In diagram (2), again either one of the two 
interaction lines can be a $\Gamma_1$, with the other line then being a
$\Gamma_2$, but here the two expressions one obtains are not identical.
Again, there is an overall symmetry factor of $2$. The same holds for
diagram (3), but without the overall symmetry factor. Diagrams (4) and
(5) can be realized only with $\Gamma_3$, and they carry no multiplication
factors. In diagrams (6) and
(7), both interaction lines must be the same, and
diagram (6) carries an extra symmetry factor of $2$. The spin structures
represent the fact that the interaction cannot flip the spin, and that
the external vertex carries a factor of $\sigma$. The functions in the
integrands of Eqs.\ (\ref{eqs:3.3}) are defined as,
\begin{mathletters}
\label{eqs:3.4}
\begin{eqnarray}
J^{(2)}(q) = \sum_k G_k\, G_{k-q}\quad,
\label{eq:3.4a}\\
J^{(3)}(q,Q) = \sum_k G_k\, G_{k-q}\, G_{k-Q}\quad,
\label{eq:3.4b}\\
J^{(4)}_1(q,Q) = \sum_k \left(G_k\right)^2\, G_{k-q}\, G_{k-Q}\quad,
\label{eq:3.4c}\\
J^{(4)}_2(q,Q) = \sum_k G_k\, G_{k-q}\, G_{k-Q}\, G_{k-q-Q}\quad,
\label{eq:3.4d}\\
I^{(2)}(q) = \sum_k G_k\, G_{-k+q}\quad,
\label{eq:3.4e}\\
I^{(3)}_1(q,Q) = \sum_k G_{-k}\, G_{k+q}\, G_{-k-Q}\quad,
\label{eq:3.4f}\\ 
I^{(3)}_2(q,Q) = \sum_k G_k\, G_{-k+q}\, G_{-k+q-Q}\quad,
\label{eq:3.4g}\\
I^{(4)}_1(q,Q) = \sum_k \left(G_k\right)^2\, G_{-k+q}\,G_{k-Q}\quad,
\label{eq:3.4h}\\
I^{(4)}_2(q,Q) = \sum_k G_k\,G_{-k+q}\,G_{k+Q}\,G_{-k+q-Q}\quad.
\label{eq:3.4i}
\end{eqnarray}
\end{mathletters}%

The information we are interested in is contained in 
Eqs. (\ref{eq:3.1}) - (\ref{eqs:3.4}) in terms of integrals. 
The remaining task is to do
these integrals. While it is easy to see by power counting that
all of the above contributions to $\chi_s$ do indeed scale like $Q^{d-1}$
for $1<d<3$, and like $O(1)$ and $O(Q^2)$ with logarithmic corrections
in $d=1$ and $d=3$, respectively,
we have found it impossible to analytically do the integrals in general,
i.e. for a finite external wavenumber in arbitrary dimensions $d$.
However, for a perturbative confirmation of the expected nonanalyticity
such a general analysis is not necessary. Rather, it is sufficient to
explicitly obtain the prefactors of the logarithmic singularities in
$d=1$ and $d=3$. If they are not zero, then by combining this with power
counting and the expected continuity of $\chi_s$ as a function of $d$, it
follows that the prefactor of the $Q^{d-1}$ nonanalyticity does not vanish
for generic values of $d$ either. For the temperature dependence at
${\bf Q}=0$ the integrals can be done in arbitrary $d$, see
Sec.\ \ref{subsec:III.E} below.

In Secs.\ \ref{subsec:III.B} - \ref{subsec:III.D} we therefore analyze 
the above integrals in $d=1$ and $d=3$. In doing so, we treat the 
particle-hole and 
particle-particle channel contributions separately, since they have quite 
different structures. We also anticipate that we will be interested only
in the static spin susceptibility, so $Q=(0,{\bf Q})$. In $d=1$, we write
${\bf Q}$ for the one-dimensional vector, i.e. a real number that can be
either positive or negative.

\subsection{$d=1$}
\label{subsec:III.B}

Let us first consider $d=1$. We do this mainly to make contact with established
results in the literature. As explained above, the particle-particle channel
must not be taken into account in $d=1$, so we put $\Gamma_3=0$. Since we
are interested in a logarithm that results from an infrared singularity, 
it suffices to calculate the integrands in the limit of small frequencies 
and wavenumbers. Be performing the integrals in Eqs.\ (\ref{eq:3.4a}) -
(\ref{eq:3.4d}) one obtains, with $Q=(0,{\bf Q})$ and $q=(\Omega_n,{\bf q})$,
\begin{mathletters}
\label{eqs:3.5}
\begin{equation}
J^{(2)}(q) = \frac{-N_F}{1 + (\Omega_n/v_F {\bf q})^2}\quad,
\label{eq:3.5a}
\end{equation}
\begin{equation}
J^{(3)}(q,Q) = N_F \left[\frac{i\Omega_n {\bf q}/{\bf Q}}{\Omega_n^2 
                   + (v_F {\bf q})^2}
              + \frac{i\Omega_n({\bf Q}-{\bf q})/{\bf Q}}{\Omega_n^2 
                   + (v_F({\bf Q}-{\bf q}))^2}\right]\ ,
\label{eq:3.5b}
\end{equation}
\begin{eqnarray}
J^{(4)}_1(q,Q) &=& N_F \left[\frac{\bf q}{\bf Q}\,\frac{(v_F {\bf q})^2 
                - \Omega_n^2} {(\Omega_n^2 + (v_F {\bf q})^2)^2}
                 - \frac{{\bf q}/{\bf Q}}{\Omega_n^2 
                    + (v_F {\bf q})^2}\right.
\nonumber\\
               &&+\frac{{\bf q}^2/{\bf Q}^2}{\Omega_n^2 
                         + (v_F {\bf q})^2}
            -\left.\frac{({\bf Q}-{\bf q})^2/{\bf Q}^2}{\Omega_n^2 
                              + (v_F({\bf Q}-{\bf q}))^2} \right]\quad,
\nonumber\\
\label{eq:3.5c}
\end{eqnarray}
\begin{eqnarray}
J^{(4)}_2(q,Q) &=& N_F \left[-\frac{2{\bf q}^2/{\bf Q}^2}{\Omega_n^2 
                                   + (v_F {\bf q})^2}
                  + \frac{({\bf q}-{\bf Q})^2/{\bf Q}^2}{\Omega_n^2 
                                 + (v_F({\bf q}-{\bf Q}))^2}\right.
\nonumber\\
               &&+\left.\frac{({\bf q}+{\bf Q})^2/{\bf Q}^2}{\Omega_n^2 
                               + (v_F({\bf q}+{\bf Q}))^2}\right]\quad.
\label{eq:3.5d}
\end{eqnarray}
\end{mathletters}%
Inserting this into Eqs.\ (\ref{eqs:3.3}), doing the final integrals, and
collecting the results one obtains, apart from analytic terms,
\begin{equation}
\chi_s({\bf Q}) = 2N_F - 4N_F(\Gamma_t N_F)^2 \ln (2k_F/\vert{\bf Q}\vert)\quad.
\label{eq:3.6}
\end{equation}
This result agrees with the well known one to this order in
$\Gamma_t$.\cite{DzyaloLarkin} One would expect that the 
$\ln\vert{\bf Q}\vert$ gets replaced by a $\ln\Omega$ or $\ln T$ if one 
works at ${\bf Q}=0$ and finite
$\Omega$ or $T$, respectively. Explicit calculations confirm this.
Of course the physical content of this perturbative result is limited,
since the ground state is not a Fermi liquid.\cite{Schulz} For later
reference we also mention that, to logarithmic accuracy, it is not
necessary to keep ${\bf Q}$ nonzero in the above calculation. If one works
at ${\bf Q}=0$ and determines the prefactor of the resulting logarithmic
divergence, then one obtains the same result as above.

\subsection{Particle-hole channel in $d=3$}
\label{subsec:III.C}

In $d=3$, both the particle-hole and the particle-particle channel
contribute to the terms we are interested in. Since the structures of
the integrals in the two channels are quite different, we first
consider the particle-particle channel. In $d=3$, the logarithm appears
only at $O({\bf Q}^2)$. Keeping ${\bf Q}$ explicitly in the integrals 
to that order
would be hard. However, as was pointed out in the preceding subsection,
to logarithmic accuracy this is not necessary. Rather, we can just
expand in ${\bf Q}$. The prefactor of the ${\bf Q}^2$ term will then be 
logarithmically
divergent, and the prefactor of the divergence will be the same as that
of the ${\bf Q}^2\ln\vert{\bf Q}\vert$ term whose presence the divergence 
signalizes. By
expanding Eqs.\ (\ref{eq:3.4b}) - (\ref{eq:3.4d}) to $O({\bf Q}^2)$, 
and dropping
the uninteresting contribution to the homogeneous $\chi_s$, we can
express all logarithmic contributions to $\chi_s$ in terms of two
integrals,
\begin{mathletters}
\label{eqs:3.7}
\begin{eqnarray}
J_1&=&{\sum_q}^{\prime}\sum_k\left(\frac{{\bf k}\cdot{\bf \hat Q}}{m}\right)^2
       \ \left(G_{k+q}\right)^5\,G_k \sum_p G_p\,G_{p-q}
\nonumber\\
    && = \left(\frac{N_F v_F}{24}\right)^2 \sum_{\bf q} 
                               \frac{1}{(v_F \vert{\bf q}\vert)^3} \quad,
\label{eq:3.7a}
\end{eqnarray}
\begin{eqnarray}
J_2 &=& \frac{1}{4} {\sum_q}^{\prime} \sum_k 
               \left(\frac{{\bf k}\cdot{\bf \hat Q}}{m}
         \right)^2\ \left(G_k\right)^4\,G_{k-q} \sum_p \left(G_p\right)^2\,
                                                                   G_{p+q}
\nonumber\\
    && = - J_1\quad,
\label{eq:3.7b}
\end{eqnarray}
\end{mathletters}%
where we have kept only the most divergent term. We find,
\begin{mathletters}
\label{eqs:3.8}
\begin{equation}
\chi^{(1)} = -8\Gamma_1\Gamma_2\ {\bf Q}^2\ J_1\quad,
\label{eq:3.8a}
\end{equation}
\begin{equation}
\chi^{(2)} = -4\Gamma_1\Gamma_2\ {\bf Q}^2\ (J_1 + J_2) =0\quad,
\label{eq:3.8b}
\end{equation}
\begin{equation}
\chi^{(3)} = -8\Gamma_1\Gamma_2\ {\bf Q}^2\ J_2\quad,
\label{eq:3.8c}
\end{equation}
\begin{equation}
\chi^{(6)} = 8(\Gamma_1^2 + \Gamma_2^2)\ {\bf Q}^2\ J_1\quad,
\label{eq:3.8d}
\end{equation}
\begin{equation}
\chi^{(7)} = 8\ {\bf Q}^2\ (-\Gamma_1^2\ J_1 - \Gamma_2^2\ J_2)\quad.
\label{eq:3.8e}
\end{equation}
\end{mathletters}%
Here we have used the fact that the structure $(J^{(3)})^2$ that appears
in $\chi^{(2)}$, $\chi^{(3)}$, and $\chi^{(7)}$, if expanded to order
${\bf Q}^2$, yields two terms, one of which gets canceled by parts of the
other. The remaining contribution can be expressed in terms of $J_2$.

We see that in the skeleton diagrams, $\chi^{(1)}$ - $\chi^{(3)}$, self
energy contributions and vertex corrections cancel each other. However,
in the insertion diagrams, $\chi^{(6)}$ and $\chi^{(7)}$, the same
cancellation is effective only in the spin singlet channel, while in the
spin triplet channel the two diagrams add up. Interpreting the logarithmic
divergence in $J_1$ as a $\ln\vert{\bf Q}\vert$ as explained above, 
we obtain for the
particle-hole channel contribution to $\chi_s$,
\begin{equation}
\chi_s^{p-h} = 2N_F + 2N_F (\Gamma_t N_F)^2 \frac{4}{9}\,
         \left(\frac{{\bf Q}}{2k_F}\right)^2\,\ln (2k_F/\vert{\bf Q}\vert)\quad.
\label{eq:3.9}
\end{equation}

\subsection{Particle-particle channel in $d=3$}
\label{subsec:III.D}

We now turn our attention to the particle-particle channel. As can be seen
from Sec.\ \ref{subsec:II.A}, diagrams (4) - (7) in Fig.\ \ref{fig:4}
contribute. From Eqs.\ (\ref{eq:3.3d}) and (\ref{eq:3.3g}) it follows that
the particle-particle channel contributions of diagrams (4) and (7) cancel
each other, so we are left with $\chi^{(5)}$ and $\chi^{(6)}$. Expanding
the functions $I^{(4)}_1$ and $I^{(4)}_2$, Eqs.\ (\ref{eq:3.4h}),
(\ref{eq:3.4i}), to order ${\bf Q}^2$ and doing the integrals, one finds that
the leading logarithmic contributions to both $\chi^{(5)}$ and $\chi^{(6)}$
can be expressed in terms of a single integral,
\begin{equation}
I = {\sum_q}^{\prime}\sum_k\left(\frac{{\bf k}\cdot{\bf \hat Q}}{m}\right)^2
       \ \left(G_{k}\right)^5\,G_{-k+q} \sum_p G_p\,G_{-p+q}\quad.
\label{eq:3.10}
\end{equation}
Inspection of the integrand shows that the leading divergency in $I$ is
a logarithm squared, in contrast to the particle-hole channel, where the
leading term is a simple logarithm. The reason is that
$\sum_p G_p\,G_{-p+q}$ contains a term $\sim \ln\vert{\bf q}\vert$ 
for ${\bf q}\rightarrow 0$,
which is just the usual BCS-type logarithm that is characteristic of
the particle-particle channel. It also depends on an ultraviolet cutoff,
since $\sum_p G_p\,G_{-p+q}$ does not exist in $d=3$ if the integration
is extended to infinity. In conjunction with the other factor in the
integrand of $I$, which is an algebraic function, this gives the
leading behavior
\begin{equation}
I \sim \int d{\bf q}\,\ln \vert{\bf q}\vert \int_0^{\infty} d\omega\ 
        \frac{{\bf q}^2 - 3(\omega/v_F)^2}
        {\bigl[{\bf q}^2 + (\omega/v_F)^2\bigr]^3}\quad.
\label{eq:3.11}
\end{equation}
While this diverges like $(\ln 0)^2$ by power counting, the prefactor of the
divergency turns out to be zero since the frequency integral in
Eq.\ (\ref{eq:3.11}) vanishes. This leads to the following conclusion for
the particle-particle channel contribution to $\chi_s$,
\begin{eqnarray}
\chi_s^{p-p} = 2N_F &+& 2N_F(\Gamma_c N_F)^2\,
                       \Bigl[0\times\left(\ln(2k_F/\vert{\bf Q}\vert)\right)^2
\nonumber\\ 
               &&+ O\left(\ln(2k_F/\vert{\bf Q}\vert)\right)\Bigr]\quad.
\label{eq:3.12}
\end{eqnarray}

Our method of expanding in powers of ${\bf Q}$, and extracting the
prefactor of the ensuing singularity, works only for the {\em leading}
nonanalytic contribution. With this method, therefore, the result
that is expressed in Eq.\ (\ref{eq:3.12}) is all we can achieve.
In order to determine the prefactor of the next-leading 
$\ln \vert{\bf Q}\vert$ term,
one would have to keep a nonzero external wavenumber explicitly. As
pointed out before in the context of the particle-particle channel,
this would be very difficult. However, for our purposes this is not
really necessary. We know that the interaction amplitudes in the
particle-hole and particle-particle channels, respectively, are
independent. Therefore, the particle-particle channel contribution 
cannot in general cancel the nonzero contribution from the particle-hole
channel that we found in Sec.\ \ref{subsec:III.C}. What we have
established is that the particle-particle channel is not more singular
than the particle-hole channel,
and for showing that the leading nonanalyticity in 
$\chi_s$ is $\ln\vert{\bf Q}\vert$
with a nonzero prefactor this is sufficient.

It should be pointed out that low-order perturbation theory probably
overestimates the importance of the particle-particle channel. Usually,
singularities in the particle-particle channel are logarithmically weaker
than those in the particle-hole channel, since a BCS-type ladder resummation
changes a $\ln x$ singularity into a $\ln\ln x$, and a $x^y$ singularity
into a $x^y/\ln x$. We expect this mechanism to work in the present
problem, so the particle-particle channel singularities are probably in fact 
asymptotically negligible compared to the particle-hole channel ones.
We also note that so far we have not really established that higher order
terms in the perturbation expansion cannot lead to stronger singularities
than the ones we found at second order in the interaction amplitudes.
This point will be further discussed in Section \ref{sec:IV} below.

\subsection{Temperature dependence of $\chi_s({\bf Q}=0)$}
\label{subsec:III.E}

In the last two subsections we have established that $\chi_s$ in $d=3$ at $T=0$
does indeed have a nonanalytic contribution proportional to 
${\bf Q}^2\ln\vert{\bf Q}\vert$.
As we pointed out in the introduction, in a Fermi liquid the wavenumber
scales like frequency or temperature, and one would therefore naively
expect a $T^2\ln T$ contribution to the homogeneous $\chi_s$ at $T>0$.
This raises the question of whether our results are compatible with those
of Carneiro and Pethick,\cite{CarneiroPethick} who did not find such a
contribution. In order to clarify this, let us calculate 
$\chi_s({\bf Q}=0,T)$ explicitly within our formalism. 
For the reasons explained in Sec.\ \ref{subsec:III.D} we restrict ourselves 
to the particle-hole channel, as did Ref.\ \onlinecite{CarneiroPethick}.

To this end, we put ${\bf Q}=0$ in Eqs.\ (\ref{eq:3.4a}) - (\ref{eq:3.4d}),
and consider the temperature dependence of $\chi^{(1)}$ - $\chi^{(3)}$,
$\chi^{(6)}$, and $\chi^{(7)}$. The relevant integrals are of the
structure,
\begin{equation}
\int d{\bf q}\,{\bf q}^2\ T\sum_{i\Omega_n} f({\bf q},i\Omega_n)\,
                                       g({\bf q},i\Omega_n)\quad,
\label{eq:3.13}
\end{equation}
which are most conveniently done by using the spectral representation for
the causal functions $f({\bf q},i\Omega_n)$ and 
$g({\bf q},i\Omega_n)$.\cite{FetterWalecka} 
Simple considerations show that there is no $T^2\ln T$ term if both $f$ and
$g$ are algebraic functions; only if at least one of them possesses a branch
cut can such a nonanalyticity arise. This immediately rules out $\chi^{(3)}$,
and the first and second contribution to $\chi^{(2)}$ and $\chi^{(7)}$,
respectively, as sources for a $T^2\ln T$. The reason is that an explicit
calculation of $J^{(3)}(q,Q=0)$, Eq.\ (\ref{eq:3.4b}), in the limit of small
$q$ shows that the only singularities in this function are poles.
The same is true for $J^{(4)}_1(q,Q=0)$ and $J^{(4)}_2(q,Q=0)$, but 
$J^{(2)}(q)$, which is minus the Lindhard function, has a branch cut,
and so all of the remaining terms potentially go like $T^2\ln T$.

Since again we are aiming only at logarithmic accuracy, we can replace
$J^{(4)}_1(q,Q=0)$ and $J^{(4)}_2(q,Q=0)$ by low-frequency, long-wavelength
expressions for which $J^{(4)}_2(q,Q=0) = -2J^{(4)}_1(q,Q=0)$. The
contributions from $\chi^{(1)}$ and $\chi^{(2)}$ therefore cancel
(remember that diagrams (1) and (2) in Fig.\ \ref{fig:4} carry multiplication
factors 4 and 2, respectively). The contributions from $\chi^{(6)}$ and
$\chi^{(7)}$ can both be expressed in terms of an integral
\begin{equation}
J = \int d{\bf q}\,{\bf q}^2\ T \sum_{i\Omega_n} J^{(4)}_1(q,Q=0)\,J^{(2)}(q)
                                                                   \quad.
\label{eq:3.14}
\end{equation}
In doing this integral one may encounter individual terms that
go like $T^2\ln T$, but all of those terms cancel,
and the leading $T$-dependence of $J$ is $T^2$. There hence is
{\em no} $T^2\ln T$ contribution to $\chi_s$ in $d=3$.

This result agrees with the conclusion of Ref.\ \onlinecite{CarneiroPethick},
which reached it on the basis of Fermi liquid theory.
We disagree, however, with the assertion of that reference that within the
framework of microscopic perturbation theory the absence of the $T^2\ln T$
is due to cancellations between vertex corrections and self energies, and
is hence a consequence of gauge invariance. What we find instead is that,
for all diagrams in Fig.\ \ref{fig:4}, the $T^2\ln T$ terms vanish
individually. This is consistent with the result of
Ref.\ \onlinecite{BealMonod}. These authors calculated $\chi_s$ in 
paramagnon approximation, which in our language corresponds to taking
only $\chi^{(6)}$ and $\chi^{(7)}$ into account, plus infinite resummations
that contribute to higher orders in the interaction amplitudes. They
reported the absence of $T^2\ln T$ terms in their calculation, rather than
their cancellation between the two diagrams.

This absence of the expected nonanalytic $T$-dependence in $d=3$ is
somewhat accidental. This can be seen from the $1$-$d$ case, where, as
was pointed out in Sec.\ \ref{subsec:III.B}, there is a $\ln T$
contribution to the homogeneous spin susceptibility. The technical
reason is that in $d=1$,
integrands whose only singularities are poles do contribute to the
$T^2\ln T$ terms. Consequently, in $d=1$ $T$ and $Q$ are interchangeable
in the logarithmic terms, while in $d=3$ they are not. Furthermore,
the same type of integrals that lead to a $\ln T$ in $d=1$, also
contribute to a $T^{d-1}$ nonanalyticity in $1<d<3$. In these dimensions
we therefore expect to find,
\begin{equation}
\chi_s^{p-h}({\bf Q}=0) = 2N_F + 2N_F(\Gamma_t N_F)^2\,c_d\,
                              (T/4\epsilon_F)^{d-1}\quad,
\label{eq:3.15}
\end{equation}
with $c_d$ a $d$-dependent, positive number.

We also mention that the absence of a $T^2\ln T$ in the self energy 
diagrams in $d=3$ does not
contradict the presence of such a term in the specific heat coefficient.
The relation between the specific heat and the Green's function is
intricate,\cite{FetterWalecka} and the resulting integrals have a different
structure from the ones that determine $\chi_s$.

\section{Discussion}
\label{sec:IV}

\subsection{Our results in a mode-mode coupling theory context}
\label{subsec:IV.A}

In this section we have a more detailed look at the mode-mode coupling
arguments that were presented in the introduction. We also stress some
analogies between classical and quantum fluids, and discuss some
important differences between clean and disordered systems.

Let us consider four distinct systems: (1) a classical
Lorentz model (i.e. a classical particle moving in a spatially random
array of scatterers\cite{Hauge}), (2) a classical fluid,
(3) a Fermi liquid with static impurities, and (4) a clean Fermi liquid. 
These systems represent classical and quantum fluids with and without 
quenched disorder, respectively. As was
pointed out in the introduction, dynamical correlations are ultimately
responsible for all of the effects discussed in this paper. However, in
classical systems they do not manifest themselves in static equilibrium
properties, while in quantum systems they do. In order to discuss the
analogies between classical and quantum systems, let us therefore digress
and consider an equilibrium time correlation function.
A convenient choice is the current-current
correlation function, whose Fourier transform determines the frequency
dependent diffusivity $D(\Omega)$. In both of the classical systems, (1) 
and (2), this
correlation function exhibits a long-time tail, so $D(\Omega)$ is
nonanalytic at $\Omega=0$. For $\Omega\rightarrow 0$ one finds for the 
classical Lorentz model,
\begin{mathletters}
\label{eqs:4.1}
\begin{equation}
D(\Omega)/D(0) = 1 + a\,i\Omega + b\,(i\Omega)^{d/2}\quad,
\label{eq:4.1a}
\end{equation}
while for the classical real fluid one finds,
\begin{equation}
D(\Omega)/D(0) = 1 - b'\,(i\Omega)^{(d-2)/2}\quad.
\label{eq:4.1b}
\end{equation}
The coefficients $b$ and $b'$ in Eqs.\ (\ref{eqs:4.1}) are positive. The 
long-time tail in the real fluid is stronger than the one in the Lorentz
gas because the former has more soft modes. More importantly,
the static scatterers in the Lorentz gas lead to a sign of the effect
that is different from the one in the real fluid. All of these features
can be understood in terms of the number and the nature of the soft modes
in these systems.\cite{ernst} In disordered Fermi liquids\cite{R} one has,
\begin{equation}
D(\Omega)/D(0) = 1 + b''\,(i\Omega)^{(d-2)/2}\quad,
\label{eq:4.1c}
\end{equation}
\end{mathletters}%
with $b''>0$. Here the sign is the same as in the classical Lorentz model,
which is due to the quenched disorder in either system. The strength of
the long-time tail, however, is equal to that in the classical real fluid.
As mentioned in Sec.\ \ref{sec:I}, the coupling of statics and dynamics
in quantum statistical mechanics leads to a related nonanalyticity in
the static spin susceptibility of a disordered Fermi liquid, namely
\begin{equation}
\chi_s({\bf Q})/\chi_s(0) = 1 - c\, \vert{\bf Q}\vert^{d-2}\quad,
\label{eq:4.2}
\end{equation}
with $c>0$.

\begin{figure}
\epsfxsize=6.0cm
\epsfysize=4.0cm
\epsffile{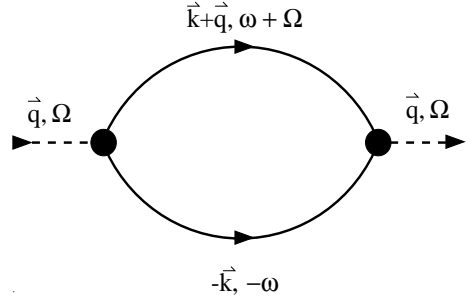}
\vskip 0.5cm
\caption{Mode-mode coupling process describing the decay of a current
 mode (dashed line) into two sound modes (solid lines).}
\label{fig:5}
\end{figure}

On the basis of these results, it is possible to predict both the strength
of the singularity, and the sign of the prefactor, in the ${\bf Q}$-dependence
of $\chi_s$ in a clean Fermi liquid, which is what we are mainly concerned 
with in this paper. In order to do so, let us recall the origin of the
nonanalyticity in the classical fluid, Eq.\ (\ref{eq:4.1b}). The density
excitation spectrum, i.e. the dynamical structure factor as measured in a
light scattering experiment, in a classcical
fluid consists of three main features: The Brillouin peaks that describe
emission and absorption of sound waves, and the Rayleigh peak that
describes heat diffusion. For our purposes, we focus on the former.
In the density-density Kubo correlation function,
$C({\bf k},\omega)$ (whose spectrum is in a classical system simply
proportional to the structure factor $S({\bf k},\omega)$), 
they manifest themselves as simple poles,\cite{Forster}
\begin{eqnarray}
C({\bf k},\omega) &\sim& \frac{1}{\omega - vk + i\gamma k^2/2}
                      + \frac{1}{\omega + vk - i\gamma k^2/2}
\nonumber\\
                  &&\equiv C_+({\bf k},\omega) + C_-({\bf k},\omega)\quad,
\label{eq:4.3}
\end{eqnarray}
where $v$ is the speed of sound, and $\gamma$ is the sound
attenuation constant. Now let us consider the simplest possible 
mode-mode coupling process that contributes to Eq.\ (\ref{eq:4.1b}),
namely one where a current mode decays into two sound modes that later
recombine, see Fig. \ref{fig:5}.
Consider a process where one of the internal sound propagators is a
$C_+$, and the other a $C_-$. At zero external wavenumber, this leads to a 
convolution integral,
\begin{mathletters}
\label{eqs:4.4}
\begin{eqnarray}
\int d\omega \int d{\bf k}\ C_+({\bf k},\omega)\ C_-(-{\bf k},-\omega + \Omega)
\nonumber\\
   \sim \int d{\bf k}\ \frac{1}{\Omega + i\gamma k^2}\sim \Omega^{(d-2)/2}\quad.
\label{eq:4.4a}
\end{eqnarray}
Note that by this mechanism the long-time tail in a system whose low-lying
modes have a linear dispersion becomes as strong as the one in a system
with diffusive modes. In contrast, if both of the sound propagators are 
$C_+$ or $C_-$, one obtains a weaker singularity,
\begin{eqnarray}
\int d\Omega \int d{\bf k}\ C_+({\bf k},\omega)\ C_+(-{\bf k},-\omega + \Omega)
\nonumber\\
  \sim \int d{\bf k}\ \frac{1}{\Omega - 2vk + i0}\sim \Omega^{(d-1)} \quad.
\label{eq:4.4b}
\end{eqnarray}
\end{mathletters}%
Now let us consider the corresponding quantum system, i.e. the clean
Fermi liquid. Again, the low-lying modes (i.e., particle-hole excitations),
have a linear dispersion. However, at zero
temperature the structure factor and the Kubo function are no longer
proportional to one another. Rather, the fluctuation dissipation theorem
shows that they are related by a Bose distribution function that eliminates
the pole at $\omega = ck$ from the structure factor. This is simply a
consequence of the fact that at zero temperature there are no excitations
that could get destroyed in a scattering process. Consequently, the
process described by Eq.\ (\ref{eq:4.4a}) is not available in this system,
and one is left with the weaker singularity of Eq.\ (\ref{eq:4.4b}).
Since the diffusion coefficient is infinite at $T=0$ in a clean system,
we look instead at the spin susceptibility as a function of ${\bf Q}$. 
${\bf Q}$ scales like $\Omega$, so we expect a singularity of the form 
$\vert {\bf Q}\vert^{d-1}$, as opposed to the $\vert{\bf Q}\vert^{d-2}$ 
in a disordered Fermi liquid, Eq.\ (\ref{eq:4.2}).
The sign of the prefactor is determined by whether or not the system
contains quenched disorder. It should therefore be opposite to the sign
in the dirty case. We thus expect for the wavenumber dependence of the
spin susceptibility in a clean Fermi liquid,
\begin{equation}
\chi_s({\bf Q})/\chi_s(0) = 1 + c'\, \vert{\bf Q}\vert^{d-1}\quad,
\label{eq:4.5}
\end{equation}
with $c'>0$. This is precisely what we found in Sec.\ \ref{sec:III} by
means of perturbation theory. Notice that the mode-mode coupling arguments
suggest that the sign of the prefactor $c'$ will be positive, regardless
of the interaction strength, as is the sign of the long-time tail in a
classical fluid. We will come back to this point in Sec.\ \ref{subsec:IV.C}
below.

\subsection{Our results in a renormalization group context}
\label{subsec:IV.B}

Another useful way to look at our results is from a renormalization group
point of view. The Fermi liquid ground state of interacting fermion
systems in $d>1$ has recently been identified with a stable fixed point
in renormalization group treatments of both a basic fermion 
theory,\cite{Shankar} and a bosonized version of that 
theory.\cite{KwonHoughtonMarston}
The instability of the Fermi liquid in $d=1$ is reflected by an infinite
number of marginal operators whose scale dimensions are proportional to
$d-1$, i.e. they all become relevant in $d<1$, and are irrelevant in $d>1$.
In the present context, the Fermi liquid nature of the ground state in $d>1$ is
reflected by the fact that the homogeneous spin susceptibility is finite in
perturbation theory. The nonanalytic corrections at finite wavenumber that
we are interested in correspond to the leading correction to scaling in the 
vicinity of the Fermi liquid fixed point, i.e. to an irrelevant 
operator with respect to that fixed point. Among the irrelevant operators,
there thus must be one whose scale dimension determines the leading
wavenumber dependence of the spin susceptibility.

An identification of this operator within the framework of a renormalization
group analysis would not only provide another derivation of our result, but
would also establish that the behavior we have found in perturbation theory
constitutes the leading ${\bf Q}$-dependence to 
{\em all} orders in the interaction
amplitudes. This program has not been carried out yet, although preliminary
results are encouraging.\cite{ustbp} This will provide a connection between
the mode-mode coupling arguments presented in the previous subsection and
renormalization group arguments that will be analogous to a corresponding
connection in classical fluids that has been known to exist for some
time.\cite{ForsterNelsonStephen}

In this context it should also be mentioned that there is no universal 
agreement that the
ground state of a weakly interacting Fermi system in $d>1$ is a Fermi liquid.
It has been proposed that there exists a relevant operator that makes
the Fermi liquid fixed point unstable, and leads to a non-Fermi liquid
ground state.\cite{Anderson} In order to destroy the Fermi liquid in $d$
dimensions, this would require a long-range effective 
interaction that falls off more slowly than $1/r^d$ at large distances.
While we do find an effective long-range interaction between the spin
degrees of freedom, it falls off like $1/r^{2d-1}$, and hence leaves the
Fermi liquid fixed point intact. The same conclusion was reached in
Ref.\ \onlinecite{CoffeyBedell} from studying the specific heat in $d=2$.

\subsection{Summary, and physical consequences of our result}
\label{subsec:IV.C}

We finally turn to a summary of our results, and to a discussion of their
physical consequences. By means of explicit perturbative calculations to
second order in the interaction,
we have found that the wavenumber dependent spin susceptibility in $d=3$
has the form,
\begin{mathletters}
\label{eqs:4.6}
\begin{eqnarray}
\chi_s({\bf Q})/\chi_s({\bf Q}=0) &=& 1 + c_3\,({\bf Q}/2k_F)^2 
                     \ln(2k_F/\vert{\bf Q}\vert)
\nonumber\\
                    && + O({\bf Q}^2)\quad.
\label{eq:4.6a}
\end{eqnarray}
We have calculated the particle-hole channel contribution to the
constant $c_3$, and have found it to be positive. More generally,
it follows from our analysis that in $d$-dimensional systems, the
spin susceptibility has a nonanalyticity of the form
\begin{equation}
\chi_s({\bf Q})/\chi_s({\bf Q}=0) = 1 + c_d\,(\vert{\bf Q}\vert/2k_F)^{d-1} 
                                     + O({\bf Q}^2)\quad,
\label{eq:4.6b}
\end{equation}
\end{mathletters}%
where the particle-hole channel contribution to $c_d$ is again positive.

A very remarkable feature of Eqs.\ (\ref{eqs:4.6}) is the sign of the 
leading ${\bf Q}$-dependence: For $d\leq 3$, $\chi_s$
{\em increases} with increasing $\vert{\bf Q}\vert$ like
$\vert{\bf Q}\vert^{d-1}$. For any physical system for which
this were the true asymptotic behavior at small ${\bf Q}$, this would have
remarkable consequences for the zero-temperature phase transition from
the paramagnetic to the ferromagnetic state as a function of the exchange
coupling. One possibility is that the
ground state of the system will not be
ferromagnetic, irrespective of the strength of the spin triplet interaction,
since the functional form of $\chi_s$ leads to the instability of any
homogeneously magnetized ground state.\cite{fluctuationsfootnote}
Instead, with increasing interaction
strength, the system would undergo a transition from a paramagnetic Fermi
liquid to some other type of magnetically ordered state, most likely a spin
density wave. While there
seems to be no observational evidence for this, let us point out that
in $d=3$ the effect is only logarithmic, and would hence manifest itself
only as a phase transition at exponentially small temperatures, and
exponentially large length scales, that might
well be unobservable. For $d\leq 2$, on the other hand, there is no
long-range Heisenberg ferromagnetic order at finite temperatures, and
the suggestion seems less exotic. Furthermore, any finite concentration
of quenched impurities will reverse the sign of the leading nonanalyticity,
and thus make a ferromagnetic ground state possible again.

Another possibility is that the zero-temperature 
paramagnet-to-ferromagnet transition is of first order.
It has been shown in Ref.\ \onlinecite{fm_clean} that the nonanalyticity
in $\chi_s({\bf Q})$ leads to a similar nonanalyticity in the magnetic
equation of state, which takes the form
\begin{equation}
t\,m - v_d\,m^d + u\,m^3 = h\quad,
\label{eq:4.7}
\end{equation}
with $m$ the magnetization, $h$ the external magnetic field, and $u>0$ a
positive coefficient. If the soft mode mechanism discussed above is the
only mechanism that leads to nonanalyticities, then the sign of the
remaining coefficient $v$ in Eq.\ (\ref{eq:4.7}) should be the same as
that of $c_d$ in Eq.\ (\ref{eq:4.6b}), i.e. $v_d>0$. This would
imply a first order transition for $1<d<3$. In this case the length
scale that in the previous paragraph would have been attributed to a
spin density wave would instead be related to
the critical radius for nucleation at the first order phase transition.
Further work will be
necessary to decide between these possibilities.

The conclusion that there is no continuous zero temperature 
paramagnet-to-ferromagnet transition is inescapable for any system 
with a particle-hole channel
interaction that is sufficiently weak for our perturbative treatment to
be directly applicable. An important question is now
whether or not it holds more generally, for systems whose interactions are
in general not weak. There are four obvious mechanism by which the sign
of the leading ${\bf Q}$-dependence of $\chi_s$ could be switched from positive
to negative: (1) Higher order contributions could lead to a sign of $c_d$ for
realistic interaction strengths that is different from the one for weak
interactions, or (2) they might lead to a stronger singularity with a
negative prefactor that constitutes the true long-wavelength asymptotic
behavior,  or (3) the particle-particle channel contribution might 
have a negative sign that overcompensates the positive contribution from
the particle-hole channel, or (4) the higher angular momentum channels that
we neglected might lead to a different sign. At this point, neither one
of these possibilities can be ruled out mathematically. However, from a
physical point of view neither one is very likely to occur. As we have
explained in Sec.\ \ref{subsec:IV.A}, both the functional form and the
sign of the nonanalyticity found in perturbation theory are in agreement
with what one would expect on the basis of a suggestive analogy with
classical fluids. Also, the renormalization group arguments sketched in
Sec.\ \ref{subsec:IV.B} make it appear likely that Eqs.\ (\ref{eqs:4.6})
constitute the actual asymptotic small-${\bf Q}$ behavior of $\chi_s$, although
an actual renormalization group proof of this is still missing. This
makes the first two possibilities appear unlikely. The third possibility
is unappealing for two reasons. First, the effective interaction in the 
particle-particle
channel is typically much weaker than the one in the particle-hole
channel. The reason is the characteristic ladder resummation that occurs
in the particle-particle channel if one goes to higher orders in
perturbation theory. This leads to an effective interaction of the
`Coulomb pseudopotential' type that is much weaker (typically by a
factor of 5 to 10) than what low order perturbation theory seems to
suggest.\cite{AndersonMorel} Second, that same resummation weakens any
singularity (cf. the discussion at the end of Sec.\ \ref{subsec:III.D}),
which probably makes the particle-particle channel singularity subleading.
Finally, the higher angular momentum
Fermi liquid parameters are usually substantially smaller than the
ones at $l=0$, which makes possibility (4) unlikely, except possibly in
particular systems.

If the sign of the nonanalyticity is, for some reason, negative at the
coupling strength necessary for a ferromagnetic transition to occur, 
at least in some systems, then in these systems the quantum phase
transition from a paramagnet to a ferromagnet at zero temperature as a
function of the interaction strength will be a conventional continuous
quantum phase transition with an interesting critical
behavior. This is because the nonanalyticity in $\chi_s$ leads to an
effective long-range interaction between spin fluctuations, which in
turn leads to critical behavior that is not mean-field like, yet
exactly solvable. This has been discussed recently in some
detail.\cite{fm_clean}

\acknowledgments

We would like to thank Roger Haydock, Gilbert Lonzarich, and Andy Millis
for stimulating discussions.
This work was supported by the NSF under grant numbers DMR-95-10185 and 
DMR-96-32978, by the DAAD, and by the DFG under grant number Vo 659/1-1.


\begin{references}
\bibitem{fieldtheoryfootnote} We use a field theoretic language in terms of
 Grassmann valued fields and their expectation values, see, e.g.,
 Ref.\ \onlinecite{NegeleOrland}.
 Equivalently, one can formulate the same theory in terms
 of operators and time ordered Green's functions, see, e.g. 
 Ref.\ \onlinecite{FetterWalecka}.
\end{references}
\end{document}